\documentclass[12pt]{iopart}
\usepackage{epsfig}

\begin{document}
\jl{2}

\title[Quasi-1D and beyond]{Dynamics of the Bose-Einstein condensate: quasi-one-dimension and beyond}

\author{Lincoln D Carr\dag, Mary Ann Leung\ddag$\|$\ and 
William P Reinhardt\ddag\dag\footnote[3]{To
whom correspondence should be addressed.}}

\address{\dag\ Department of Physics, University of Washington,
Seattle, WA, 98195-1560, USA\\E-mail: lcarr@u.washington.edu}

\address{\ddag\ Department of Chemistry, University of Washington,
Seattle, WA, 98195-1700, USA\\E-mail: rein@chem.washington.edu}

\address{$\|$\ Mills College, Oakland, CA, 94613-1301, USA\\
E-mail: mleung@best.com}

\begin{abstract}
It is shown that the quasi-one-dimensional Bose-Einstein condensate is experimentally accessible and rich in intriguing phenomena.  We demonstrate numerically and analytically the existence, stability, and perturbation-induced dynamics of all types of stationary states of the quasi-one-dimensional nonlinear Schr\"odinger equation for both repulsive and attractive cases.  Among our results are:  the connection between stationary states and solitons; creation of vortices from such states; manipulation of such states with simple phase profiles; demonstration of the fragility of the condensate phase in response to shock; and a robust stabilization of the attractive Bose-Einstein condensate.

\end{abstract}

\pacs{03.75.Fi, 05.30.Jp, 05.45.Yv}



\section{Introduction}
\label{sec:intro}

A family of Bose-Einstein condensates (BECs) has been created in a series of ground-breaking experiments~\cite{davis1,anderson1,fried1,bradley1,bradley2} over the last five years.  These condensates, made of trapped, dilute gases of $^{23}$Na, $^{87}$Rb, $^{1}$H, or $^{7}$Li, are well-modeled by the nonlinear Schr\"odinger equation (NLSE)~\cite{dalfovo1}, also known as the Gross-Pitaevskii equation~\cite{pitaevskii1,gross1}.  However, they have yet to be experimentally investigated in quasi-one-dimension (quasi-1D.)  We define the BEC to be in the quasi-1D regime when its transverse dimensions are on the order of its healing length and its longitudinal dimension is much longer than its transverse ones.  In this case the 1D limit of the 3D NLSE is appropriate, rather than true 1D or 2D mean field theories~\cite{kolomeisky1}.

The 1D NLSE is expected to be the correct description of the quasi-1D BEC~\cite{carr15,carr16}.  It is a ubiquitous equation.  Among other natural phenomena, it models Bose-condensed photons~\cite{ciao1}, disordered media~\cite{mamaev1}, helical excitations of a vortex line~\cite{hasimoto1}, and light pulses in optical fibers~\cite{hasegawa1}.  In this latter context soliton solutions to the NLSE have been used to greatly increase the efficiency of communications~\cite{agrawal1}.  

Because the BEC is a coherent matter-wave it has a natural analogy with light.  This insight has led to light-inspired experiments, as for example four-wave mixing~\cite{deng1} and both continuous-wave and pulsed atom lasers~\cite{hagley1,mewes1}.  The creation of a quasi-1D BEC would allow the exploration of a great body of experimental and theoretical results from fiber optics~\cite{agrawal1}.

In addition to the analogy with fiber optics there are several other reasons to explore the quasi-1D BEC.  Cold atom waveguides have recently been experimentally realized~\cite{key1,dekker1}.  This work suggests the manipulation of matter waves in quasi-1D and perhaps the eventual creation of matter-wave circuits, just as fiber optics have come to replace wires in certain applications.  The quasi-1D regime of the BEC is qualitatively different from its 2D and 3D counterparts:  in the attractive case collapse is prevented, while in both the repulsive and attractive cases solitons are stablized.

In a pair of recent articles all stationary states of the 1D NLSE under box and periodic boundary conditions~\cite{carr15,carr16} were analytically elucidated.  This provides a starting point for the analysis of the quasi-1D BEC.  In the present investigation we impose box or periodic boundary conditions in the longitudinal dimension and box boundary conditions in the transverse dimensions.  These boundary conditions model the potential quasi-1D regime of a number of present experiments: the atom waveguide~\cite{key1,dekker1}; the prolate harmonic trap in which is formed a cigar-shaped BEC~\cite{andrews1,burger1,ketterle1}; the toroidal trap~\cite{close1}; and finally an oblate harmonic trap with a barrier formed in the middle either by a second spin state of the same atom~\cite{matthews1,williams1} or by a laser~\cite{ketterle1}, in which is formed a pancake-shaped BEC with the center removed.  Periodic boundary conditions are the best first model for toroidal geometries; box boundary conditions are a good starting model for cigar-shaped geometries.  

Our approach is to find the stationary states of the NLSE under simple, analytically solvable boundary conditions.  These give physical insight into the corresponding results for potentials which require numerical solution.  Here we demonstrate the merits of this approach by the extension of 1D into quasi-1D, 2D, and 3D.  Elsewhere we have applied such insight to the harmonic potential~\cite{carr19}, among others.

The paper is organized as follows.  In section~\ref{sec:stat} we review the analytic stationary solutions in 1D.  In section~\ref{sec:1D} we numerically illustrate these solutions and their properties and predict how they should extend into quasi-1D and higher dimensions.  In section~\ref{sec:q1D} we numerically verify these predictions and thereby demonstrate that the quasi-1D regime is experimentally accessible for both repulsive and attractive condensates.  In section~\ref{sec:2D3D} we show how quasi-1D-type stationary states extend into 2D and 3D and the rich phenomena that result.  Finally in section~\ref{sec:conclusion} we give our conclusions.

\section{Solitons and stationary states}
\label{sec:stat}

We consider two cases of the BEC: repulsive and attractive atomic pair interactions.  For our purposes they differ by the sign of the nonlinear term in the NLSE.  The former is the one that has received the most experimental attention.  The latter is generally unstable in 3D~\cite{ruprecht1} but stable in 1D~\cite{kivshar5}.

The NLSE in 1D has a number of special properties which are described in the mathematical literature.  It is integrable~\cite{kivshar3,sulem1}, may be solved exactly by the Inverse Scattering Transform~\cite{zakharov1,zakharov2} and has a countably infinite number of conserved quantities~\cite{miura1,miura2}.  We have taken advantage of its special properties in finding its stationary solutions.  We first describe all stationary states of the 1D NLSE subject to box and periodic boundary conditions.  We then make clear their connection with solitons and provide physical interpretations of their form.

\subsection{Mathematical form}
\label{subsec:math}

We begin with the NLSE that describes a BEC of $N$ atoms
of mass $M$, confined in an external potential $V(\vec{r})$:

\begin{equation}
[-\frac{\hbar^{2}}{2M}\nabla^{2}+g\mid\!\psi(\vec{r},t)\!\mid^{2}+V(\vec{r})\,]\,\psi(\vec{r},t) = \imath\hbar\partial_{t}\,\psi(\vec{r},t)
\label{eqn:nlse3D}
\end{equation}
where $\mid\!\!\psi(\vec{r},t)\!\!\mid^{2}$ is the single particle density such that $\rho(\vec{r},t)=N\mid\!\!\psi(\vec{r},t)\!\!\mid^{2}$, the coupling constant $g\equiv 4\pi\hbar^{2}aN/M$, and $a$ is the scattering length for binary collisions between atoms.  The case of repulsive interactions corresponds to $a>0$; that of attractive interactions to $a<0$. 

The characteristic length scale of variations in the condensate wavefunction is the healing length $\xi$:

\begin{equation}
\label{eqn:xidefinition} 
\xi \equiv (8\pi\bar{\rho}\mid\!a\!\mid)^{-1/2}
\end{equation}
where $\bar{\rho}$ is the mean particle density.  The BEC is in the quasi-1D regime when the lengths of the tranverse $y$ and $z$ dimensions of the trap, $L_{y}$ and $L_{z}$, satisfy the following criteria:  $L_{y},L_{z}\sim\xi_{eff}$ and $L_{y},L_{z}<\!<L_x$.  The former ensures that the condensate remains in the ground state in the two transverse dimensions while the latter ensures that longitudinal excitations are much lower in energy than possible transverse excitations.  Under these conditions one may make an adiabatic separation of longitudinal and transverse variables~\cite{carr15,perez1}.  Equation~(\ref{eqn:nlse3D}) then reduces to:

\begin{equation}
[-(\xi_{eff})^2\partial_{x}^{2}\pm\mid\!f(x)\!\mid^{2}+\tilde{V}(x)\,]\,f(x) = \imath\partial_{\tilde{t}}\,f(x)
\label{eqn:nlsetime}
\end{equation}
where all terms have been made dimensionless: $f(x)$ is the longitudinal portion of the wavefunction; $\tilde{t}\equiv(\hbar/2M\xi^{2})t$ we will call the natural time; the $\pm$ refer to the repulsive and attractive cases, respectively; $\tilde{V}(x)\equiv(2M\xi^{2}/\hbar^{2})V(x)$;  and $\xi_{eff}$ is an effective healing length, discussed in section~\ref{subsec:connect}.  If we furthermore assume a stationary state:

\begin{equation}
[-(\xi_{eff})^2\partial_{x}^{2}\pm\mid\!f(x)\!\mid^{2}+\tilde{V}(x)\,]\,f(x) = \tilde{\mu}\,f(x)
\label{eqn:nlse}
\end{equation}
where $\tilde{\mu}\equiv(2M\xi^{2}/\hbar^{2})\mu$ is a dimensionless chemical potential which is now an eigenvalue.

All stationary solutions to equation~(\ref{eqn:nlse}) may be written in terms of Jacobian elliptic functions~\cite{bowman1}.  The properties of such functions are reviewed elsewhere~\cite{carr15,bowman1,abramowitz1}.  There are five normalizable symmetry-breaking solution-types to the stationary NLSE with a constant potential on a finite interval.  They are pictured in figure~\ref{fig:5solns}.  The stationary solutions appropriate to longitudinal box boundary conditions are:

\begin{equation}
f(x)=A\,{\rm sn}(\frac{2j K(m) x}{L_x}+\delta\mid m)
\label{eqn:sn}
\end{equation}
\begin{equation}
f(x)=A\, {\rm cn}(\frac{2j K(m) x}{L_x}+\delta\mid m)
\label{eqn:cn}
\end{equation}
for the repulsive and attractive cases, respectively.  Equation~(\ref{eqn:sn}) is shown in figure~\ref{fig:5solns}(a) and equation~(\ref{eqn:cn}) is shown as the anti-symmetric solution in figure~\ref{fig:5solns}(c).  $A$ is the amplitude, $j-1$ is the number of nodes, $K(m)$ is the complete Jacobian elliptic integral which is the quarter period of these functions, $\delta$ is the offset, here equal to zero, and $0\leq m\leq 1$ is the Jacobian elliptic parameter.  The general notation sn$(u\mid m)$ is standard for Jacobian elliptic functions~\cite{bowman1,abramowitz1}.

Under periodic boundary conditions equations~(\ref{eqn:sn}) and~(\ref{eqn:cn}) are still solutions, but $j$ is required to be even and the number of nodes is $j$, i.e. $j=2,4,6,...$  Translational symmetry is restored, so the offset $\delta$ is no longer required to be zero.  This leads to a Kosterlitz-Thouless-type entropy in 1D~\cite{carr15}.

In addition to the above two solution-types there are three nodeless, symmetry-breaking solution-types under periodic boundary conditions.  The first, a solution to the attractive case, is real:

\begin{equation}
f(x)=A\, {\rm dn}(\frac{2j K(m) x}{L_x}+\delta\mid m)
\label{eqn:dn}
\end{equation}
An example is shown as the symmetric solution in figure~\ref{fig:5solns}(c).  The other two types are intrinsically complex.  $f(x)\equiv r(x)\exp(\imath\phi(x))$ and for the repulsive and attractive cases, respectively:

\begin{equation}
r(x)^{2}=A^2(1-\gamma\,{\rm dn}^{2}(\frac{2jK(m) x}{L_x} + \delta \mid m))
\label{eqn:repcom}
\end{equation}
\begin{equation}
r(x)^{2}=A^2({\rm dn}^{2}(\frac{2jK(m) x}{L_x}\mid m)-\gamma(1-m))
\label{eqn:attcom}
\end{equation}
where in both cases the phase must be found by numerical integration from the equation:

\begin{equation}
\phi^{\prime}(x)=\frac{\alpha}{r(x)^2}
\label{eqn:comphase}
\end{equation}
The phase and amplitude are shown in figure~\ref{fig:5solns}(b) and~\ref{fig:5solns}(d).  In the repulsive case $A^2\gamma$ is the depth of the density minima below the constant background.  When $\gamma = 1$ equation~(\ref{eqn:sn}) is recovered.  In the attractive case $\gamma$ interpolates between the real anti-symmetric and symmetric solutions in equations~(\ref{eqn:cn}) and~(\ref{eqn:dn}).  In both cases $0\leq\gamma\leq1$.  $\alpha$ is a constant of integration.  $j$ is the number of density minima or maxima, respectively.

All five solution-types are found by solving equation~(\ref{eqn:nlse}) subject to normalization and boundary conditions, and are described in detail in references~\cite{carr15} and~\cite{carr16}.

\subsection{Physical interpretation}
\label{subsec:interpret}

The nonlinear mean field term in equation~(\ref{eqn:nlse}) causes a spreading out or a clumping together of the wavefunction between nodes for the repulsive and attractive cases, respectively.  Thus the stationary solutions to the NLSE with box boundary conditions are in one-to-one correspondence with the sinusoidal solutions to the particle-in-a-box problem in linear quantum mechanics.  The number of nodes is $j-1$ and the complete set of stationary states is described by equations~(\ref{eqn:sn}) and~(\ref{eqn:cn}), where $j\in\{1,2,3,...\}$.

Likewise, if $j$ is restricted such that $j\in\{2,4,6,...\}$, equations~(\ref{eqn:sn}) and~(\ref{eqn:cn}) are in one-to-one correspondence with the sinusoidal solutions to the particle-on-a-ring problem in linear quantum mechanics.  As the ring has rotational symmetry these solutions are symmetry-breaking.  We note that, as in the linear Schr\"odinger equation, there are also complex, constant-amplitude, plane-wave solutions on the ring.

The Jacobian elliptic parameter $m$ governs the strength of the nonlinearity.  As $m\rightarrow 0^+$, sn $\rightarrow$ sin and cn $\rightarrow$ cos, respectively.  This is the linear, sinusoidal limit.  As $m\rightarrow 1^-$, sn $\rightarrow$ tanh and cn $\rightarrow$ sech.  These are the dark and bright soliton, stationary solutions to the NLSE on the infinite line, respectively~\cite{zakharov1,zakharov2}.  This shows the connection between these stationary solutions and solitons.  In the box the $j^{{\rm th}}$ stationary state is a $j-1$ or $j$ soliton-train in the repulsive and attractive cases, while on the ring the $2j^{{\rm th}}$ stationary state is a $2j$ soliton-train.  As we have shown elsewhere~\cite{carr19}, soliton-trains interact in a manner similiar to single solitons.  And as with single solitons their form and phase relationships are quite robust.

The other three solution-types have no analogue in the particle-on-a-ring problem in linear quantum mechanics.  They exist only on the ring and are nodeless and symmetry-breaking.  The complex solutions have a monotonically increasing phase.  Each has a complex-conjugate, degenerate partner.  The complex solutions for the attractive case, in particular, are an entirely new stationary solution-type for the NLSE.

The complex solutions for the repulsive case are interpreted as density-notch solitons moving with speed $c$ on the ring with an opposing momentum boost of the condensate of speed $-c$, which results in a stationary state in the lab frame.  Density-notch solitons have a speed between zero and the Boguliubov sound speed, ranging from maximal to zero depth, respectively~\cite{reinhardt1}.  Those not of maximal depth are called grey solitons, while those which are of maximal depth and therefore form a node are called dark.  Figure~\ref{fig:5solns}(b) shows the bounded, quantized version of a grey 2-soliton-train.

At typical experimental trap sizes the single density-notch stationary state on the ring is the lowest energy, symmetry-breaking excitation above the real, constant-amplitude ground state.  When any of these stationary excited states is perturbed it gives rise to soliton-type motions~\cite{reinhardt1}.  Recent reports suggest that such motions can be induced in repulsive BECs by optical phase engineering of the condensate phase~\cite{burger1,denschlag1}.

All attractive symmetry-breaking, longitudinally periodic, stationary solution-types, i.e. the anti-symmetric and symmetric ones shown in figure~\ref{fig:5solns}(b) and the complex one shown in figure~\ref{fig:5solns}(d), are described by the $C_j$ point symmetry group, where $j$ is the number of peaks.  There are $j$ nearly degenerate solutions.  For even $j$ there is a real symmetric-anti-symmetric pair and $(j-2)/2$ degenerate complex pairs.  For odd $j$ there is a real symmetric solution and $(j-1)/2$ degenerate complex pairs.  Thus by group theory these three solution-types form the complete set of stationary states made of evenly spaced peaks.

As the attractive BEC has been made in 3D but not in quasi-1D, these stationary solutions are open to experimental investigation.

\subsection{Phase engineering of dynamics}
\label{subsec:1Dmanipulation}

There are two ways to impart velocity to a density-notch soliton.  One may boost the constant background via a \emph{phase ramp} and the notch will drift with the resulting super-current.  One may also apply a \emph{phase jump} and the resulting soliton speed with respect to the background will be $c=c_{max}\sin(\delta/2)$, where $c_{max}$ is the Boguliubov sound speed and $\delta$ is related to the phase difference across the moving density notch~\cite{reinhardt1}.  Both types of phase profile are apparent in figure~\ref{fig:5solns}(b).  The phase ramp may be seen in the constant background slope in the phase.  The phase jump is shown over each of the two notches.  The velocities cancel and a stationary state results.

To impart a collective velocity to trains of density-notches one applies an equal phase difference across each of the member notches or a phase ramp across the whole train.  One may also treat the members of the train individually by applying varying jumps and thereby study solitons interactions.  A bright soliton may be set in motion by a phase ramp but not a phase jump.  For a phase ramp $\exp(\imath k x)$ the resulting velocity $c\propto k$.  There is no upper bound on the velocity of bright solitons.  As with density-notches, a train of bright solitons may be manipulated collectively by applying a phase ramp across the whole set or individually by applying different phase ramps across the members of the train.

Soliton-trains are thus collective excitations which can be directly manipulated by simple phase profiles.  One may manipulate the members of the train individually or collectively.  This makes clear the connection between the above stationary states and solitons.

\section{One dimension}
\label{sec:1D}

\subsection{Stability}
\label{subsec:1Dstability}

We first demonstrate that the five solutions types described in section~\ref{sec:stat} are indeed stationary.  In figure~\ref{fig:1D} we propagate them in time numerically.  In the numerical work presented for one, two, and three dimensional solutions of the NLSE the condensate wave function was expanded as $\psi(\vec{r},t)=\Sigma_i c_i(t)\phi_i(\vec{r})$.  The $\phi_i(\vec{r})$ were chosen as a co-ordinate-discretized, pseudo-spectral basis, and the $c_i(t)$ were propagated in time using a fourth order variable-step Runge-Kutta algorithm, made highly efficient by implementation with spatial fast fourier transform techniques.  We shall use the same numerical methods throughout our article.

The denumerably infinite number of conserved quantities in the 1D NLSE makes solitons robust.  Single solitons, in whatever equations they arise, are in general stable.  Here in both the repulsive and attractive cases soliton-trains are stable~\cite{carr19}.  In the attractive case the two nodeless solution-types are \emph{cyclically stable}.  That is, in response to stochastic perturbation they evolve in a complicated manner but come asymptotically close to their original state in an aperiodic cycle.  In figure~\ref{fig:1Dnoise} we illustrate the stability properties of the stationary states by introducing $1\%$ stochastic white noise into all five fundamental stationary solution types and observing their evolution.

Stochastic perturbation of density-notch solitons can cause a velocity shift or small-scale emission of phonons~\cite{kivshar1}.  The phase relationships between solitons, however, remain intact.  The depth of a density-notch is a simple function of its velocity.  Therefore the soliton-train changes in depth but not in spacing or form.  It drifts as a unit in a diffusive manner.  This is true for both the real and complex solution-types.

Stochastic perturbation of bright solitons can cause a velocity shift, small-scale emission of phonons, and/or a phase shift~\cite{elgin1}.  Although a single soliton is stable, multiple solitons are cyclically stable.  As the phase difference between adjacent solitons changes from zero to $\pi$ their interaction changes from attractive to repulsive~\cite{carr19}, and it is this changing interaction that leads to cyclical stability.  The anti-symmetric solution is stable.  The repulsive interactions between solitons lock them into place.  In the other two solution-types the solitons attract and repel as their relative phases are perturbed by the phonons that make up the white noise.  We have not shown their recurrence here.

\subsection{Expectations in higher dimensionality}
\label{subsec:predict}

We expect that in quasi-1D the five solution-types described analytically in section~\ref{sec:stat} and numerically in section~\ref{subsec:1Dstability} will have the same properties as in 1D.  We also expect, and will verify in the following section, that excellent quasi-1D stationary states can be found as the direct product of eigenstates of the 1D NLSE in each of the dimensions, even though the NLSE is only approximately separable.

To reiterate our criteria for quasi-1D from section~\ref{subsec:math}: $L_{y},L_{z}\sim\xi_{eff}$; $L_{y},L_{z}<\!<L_x$.  In the repulsive case a non-overlapping 2-soliton solution becomes possible when the length is $2\pi\xi_{eff}$~\cite{carr15}.  This provides a criterion for the breakdown of the quasi-1D approximation: if $L_y$, $L_z > 2\pi\xi_{eff}$ a transverse excitation becomes likely.

Solutions to the attractive NLSE become unstable when the negative mean field energy is not sufficiently opposed by the positive kinetic energy.  This may be seen by performing a variational calculation.  So one criterion we could use is the length for which the transverse solution changes from positive to negative energy.  This is about $2 \frac{1}{2} \xi_{eff}$.  In the literature it has been found that the 3D attractive NLSE is more likely to go unstable than the 2D NLSE~\cite{berge1}.  So we expect that the quasi-1D approximation will break down at a smaller transverse length in 3D than in 2D.  In the attractive case we expect that out of the quasi-1D regime the wavefunction should collapse.

We note that in the repulsive case the criteria for separability and quasi-1D are not identical.  In the limit as the transverse lengths become very large the NLSE becomes exactly separable for a constant transverse solution.  However, the separable solutions can be unstable, as we shall show in section~\ref{sec:2D3D}.

In the repulsive case it has been shown in the mathematical literature that band soliton and planar soliton solutions of the NLSE in 2D and 3D can decay into vortices via long-wavelength transverse modulations known as snake instabilites~\cite{jones1,kivshar3}.  They are called band or planar solitons both to describe their shape in 2D and 3D and to differentiate them from 1D solitons of the type formally obtained by the Inverse Scattering Transform~\cite{zakharov1,zakharov2}.  Thus as we increase the length of the transverse dimensions out of the quasi-1D regime and into 2D and 3D we expect to find that our soliton-train solutions break up into vortices.  The BEC conserves vorticity.  When the band or planar solitons decay the circulation quantum numbers of the resulting vortices should add to zero.  Since the energy of a vortex is proportional to its quantum number squared we expect the decay to be into pairs of singly-quantized vortices of opposite circulation.

\section{Quasi-one-dimension}
\label{sec:q1D}

To demonstrate the experimental viability of the 1D stationary states described above we add one or two transverse dimensions of scale size $\xi_{eff}$ and study the stability and phase engineering of solitons and stationary states in quasi-1D.  The connection between our numerical studies and present BEC experiments is made clear.  We also illustrate the phase rigidity of the quasi-1D BEC.  We hope the following work will encourage experimentalists to create a quasi-1D attractive BEC and to study density-notches and phase rigidity in the quasi-1D regime of the repulsive BEC.

\subsection{Connection with experiments}
\label{subsec:connect}

In table~\ref{tbl:exp} we show the length and time scales and the number of atoms for the four atomic species of BEC.  The below conversion formula allows one to find the number of atoms given the healing length, trap size, and type of BEC.

\begin{equation}
N=\frac{\xi_{eff} L_x L_y L_z}{8\pi a}
\label{eqn:numatoms}
\end{equation}
where the healing length $\xi_{eff}$ and the scattering length $a$ must be in the same units and the trap dimensions $L_x$, $L_y$, and $L_z$ are in units of $\xi_{eff}$.  From section~\ref{subsec:math} the time scale is:

\begin{equation}
t=\frac{2 M \xi_{eff}^2}{\hbar}\tilde{t}
\label{eqn:timescale}
\end{equation}
where $M$ is the mass of the atomic species.  $\tilde{t}$ are the natural, unitless time units we use in the numerical studies.

In the table we choose $\xi_{eff}=3.0\,\mu$m to make solitons in the condensate easily observable.  This is the order of the healing length used in recent soliton experiments~\cite{burger1,denschlag1}.  The time scales are on the order of 0.1 to 1 ms per natural time unit.  Since we present results on time scales from 25 to 100 natural time units all our numerics take place on the experimental time scales of from 1 to 100 ms.  Similiar results persist to 1000 natural time units.  As BECs last from 1 to 75 s our studies extend over the lifetime of current BECs~\cite{dalfovo1,ketterle1}.

For computational reasons we have used somewhat smaller traps than current experiments, 10x1x1 or 10x$\frac{1}{2}$x$\frac{1}{2}$ in the attractive case and 25x1x1 in the repulsive case.  Thus the number of atoms is small.  In the attractive case the trap can be made longer and remain in the quasi-1D regime as long as the number of peaks in the solutions is increased.  In the repulsive case the trap size could be increased in each transverse dimension by a factor of 5 and arbitrarily in the longitudinal dimension, resulting in a factor of 100 or more increase in the volume.  Therefore the numbers we have used in the simulations scale into the experimentally accessible regime.

The assumption of separability in quasi-1D is made by projecting the total wavefunction onto the ground states of the transverse dimensions and integrating.  This produces an effective healing length which depends on the transverse width:

\begin{equation}
\xi^{2}_{eff}=(\frac{3\,(E(m_t)/K(m_t)-1)^2}{-2(1+m_t)E(m_t)/K(m_t)-(2+m_t)})^2\,\xi^2
\label{eqn:repeff}
\end{equation}
\begin{equation}
\xi^{2}_{eff}=(\frac{3\,(E(m_t)/K(m_t)-(1-m_t))^2}{2(-1+2m_t)E(m_t)/K(m_t)-(2-5m_t+3m_t^2)})^2\,\xi^2
\label{eqn:atteff}
\end{equation}
for the repulsive and attractive cases, respectively.  $m_t$ is the Jacobian elliptic parameter which determines the width of the transverse wavefunction, which is a sn or cn function described by equations~(\ref{eqn:sn}) or~(\ref{eqn:cn}).  $E(m_t)$ and $K(m_t)$ are complete Jacobian elliptic integrals.

In the limit as $m_t\rightarrow 0^+$, $\xi^2_{eff}\rightarrow\frac{4}{9}\xi^2$ in both cases and the transverse wavefunction is sinusoidal.  As $m_t\rightarrow 1^-$, $\xi^2_{eff}\rightarrow\xi^2$ and $\xi^2_{eff}\rightarrow 0$ monotonically in the repulsive and attractive cases, respectively.  For the purposes of the present quasi-1D numerical investigations we shall use a transverse healing length of $1$ or $\frac{1}{2}$. For these healing lengths $m_t\sim 0.1$ and $\xi^{2}_{eff}\sim\frac{4}{9}\xi^{2}$.  We have used these rescalings in determining the number of atoms in table~\ref{tbl:exp}.

Besides the ratio of the box lengths to the healing length, the numerical studies do not depend on the choice of the experimental parameters.  In all figures we have set $\xi_{eff}=1$ and thus are working in length units scaled to the effective healing length.

\subsection{Stability}
\label{subsec:q1Dstability}

As in section~\ref{subsec:1Dstability} we again add $1\%$ stochastic white noise to the five solution-types, but in both the 2D and 3D quasi-1D limits.  The initial state is created from the product of box transverse wavefunctions of the form described by equations~(\ref{eqn:sn}) and~(\ref{eqn:cn}) with a box or periodic longitudinal wavefunction of each of the five types shown in figure~\ref{fig:5solns}.

In figure~\ref{fig:quasi1D} we show results of simulations in the 3D case.  Trap parameters which enforce the quasi-1D dynamics are used: a volume of 25x1x1 in the repulsive case and 10x$\frac{1}{2}$x$\frac{1}{2}$ in the attractive case.  The stability properties are identical with those found in 1D.  Although we have not shown it here, the 2D case displays essentially the same results.

The repulsive solutions drift but are otherwise unchanged.  The complex solution-type moves with a steady velocity to the left because the approximation of separability has failed to give the right momentum boost of the condensate to counteract the motion of the soliton-train.  But clearly a larger momentum boost would return it to a stationary state.  Note that the phase relationships are quite stable: constant between notches in figure~\ref{fig:quasi1D}(a) and changing linearly between notches in figure~\ref{fig:quasi1D}(c).

The attractive anti-symmetric solution is locked into place and is stable as it was in 1D.  The two nodeless solution-types break up and recur.  The symmetric, real type shown in figure~\ref{fig:quasi1D}(g)-(h) shows an example of this recurrence.  In the third time slice the left-hand peak has spread out and changed its phase relationship to the right-hand peak.  In the fourth time slice it has returned to its initial state.

As predicted, there are small fluctuations due to the assumption of separability in our choice of initial state.  In the repulsive case we found that beyond six healing lengths the solutions became unstable in both 2D and 3D.  In the attractive case we found that the requirements for the quasi-1D regime were more complicated.  A transverse length of $\frac{1}{2}\xi_{eff}$ was required in 3D while $\xi_{eff}$ was sufficient in 2D.  In general the stability depended both on the number of peaks and the solution-type.  The anti-symmetric solution-type with a large number of peaks was the most stable configuration.  This kept the density spread out and the contribution of the mean field term in the NLSE small.  Other researchers have found similiar results~\cite{michinel1}.

\subsection{Manipulation via phase}
\label{subsec:q1Dmanip}

In section~\ref{subsec:1Dmanipulation} we discussed simple phase profiles which impart a velocity to soliton-trains.  Setting the whole train in motion around a thin torus creates a stable, structured super-current in both repulsive and attractive cases.  Measuring the velocity of the train is a way to verify the superfluidity of the gaseous BEC.  One method would be combination of the new toroidal trap, created to Bose condense Cs~\cite{close1}, with optical phase-imprinting~\cite{dobrek1}.  Note that the superfluid properties of the attractive BEC are presently unknown.

One may study soliton interactions by treating members of the train separately.  In figure~\ref{fig:phaseManip} we show that planar solitons in quasi-1D interact in a manner similiar to solitons in 1D.  In the repulsive case we have used an 0.3$\pi$ phase jump to send two density notches towards one another.  As may be seen in the figure, they undergo elastic repulsive interaction twice over 50 natural time units.

In the attractive case we have used a real symmetric initial state with a phase ramp of 2$\pi$ on each bright soliton.  Their first and third interactions show a characteristic smearing of the density which indicates a phase difference between zero and $\pi$~\cite{carr19}.  In the second interaction they add coherently as they pass through each other, indicating a phase difference of zero.  In their fourth interaction they repel.  Therefore it is quite apparent that their relative phase has drifted by approximately $\pi$, due to the $1\%$ stochastic noise added to the initial state.

For computational reasons we chose phase profiles which caused interaction on a short time scale.  But similar interactions could be observed on longer time scales by the appropriate choice of phase jump or ramp.  Therefore solitons and stationary states in quasi-1D are connected in the same way as in 1D.  Soliton interactions can be studied experimentally in the BEC.

By use of phase ramps one may also study how the BEC responds to shock.  The repulsive case has the property of \emph{phase rigidity}, related to symmetry-breaking of the condensate phase.  Rigidity is common to many symmetry-breaking materials in condensed matter physics, from liquid crystal to chalk~\cite{anderson2}.  The BEC exhibits this property by shattering into well-defined phase domains, the size of which are on the order of the healing length.  In quasi-1D the borders between domains are made of density-notch solitons.

To illustrate this property we use longitudinal box rather than periodic boundary conditions, with volume 25x1x1.  A large phase ramp of $\phi(x)=1.6\pi x$ is applied to a two-density-notch stationary state.  In figure~\ref{fig:shock1D} we show the time evolution of such a configuration.  The first time slice shows the initial state.  The second shows a shock wave, apparent in the density.  The final time slice shows a highly excited state with well-defined phase domains separated by density notches.  Because the kinetic energy of this state is very high the nonlinear term in the NLSE does not contribute signifigantly.  Thus a linear beat phenomenon is seen in the density.  We have used short time scales for convenience, but a smaller phase ramp would cause the same phenomenon to occur on a much longer, experimentally observable time scale.

\section{Beyond quasi-one-dimension}
\label{sec:2D3D}

In this section the transverse dimensions are extended beyond the quasi-1D regime in both two and three dimensions.  Band and planar solitons become unstable, and shocking the condensate produces chaos.

\subsection{Vortex creation}
\label{subsec:vortex}

In figures~\ref{fig:vortex2D} and~\ref{fig:vortex3D} we show the decay of band and planar solitons into vortices.  In both cases we begin with 2-soliton-train stationary states made in the same way as described in section~\ref{sec:q1D}.  We use dimensions 25x12 in 2D and 25x12x12 in 3D with periodic boundary conditions in the longitudinal direction.  White noise causes the solitons to decay into a pair of vortices of opposite circulation, conserving vorticity.  The vortices are singly quantized, as may be seen in the figures by traversing a node in the phase.  The colors pass once around the phase color circle, which is defined to be 2$\pi$ in extent.  In the plots we chose times that would highlight the decay.  However, the vortices are stable, spatially stationary, and do not interact over the full length of the study.

White noise in the BEC is principally due to uncondensed atoms and the number of such atoms is a function of temperature.  Thus noise is nominally an experimental parameter.  The time from creation of the initial stationary solitons to their decay into vortices is a function of the noise.  With the correct choice of temperature and volume, stationary band or planar solitons, made via phase engineering~\cite{denschlag1}, can be used to create vortices in a controlled manner.

\subsection{Vortex interaction}
\label{subsec:interact}

Band and planar solitons respond to simple phase profiles in the same way as solitons in quasi-1D.  The following method permits direct study of vortex interactions.

As explained above, the time from creation of the initial state to decay into vortices is a function of the temperature of the BEC.  As in quasi-1D, they move toward each other with a velocity which depends on the phase jump.  Thus two solitons can be made to decay into vortices shortly before collision.  An example of this process in 2D is shown in figure~\ref{fig:interact}.  We use a phase jump of $\pi$/15, a noise level of $0.1\%$, and a container identical to the one in figure~\ref{fig:vortex2D}.  In the first time slice a snake instability and a vortex pair have developed.  By $t=36$ the vortices are strongly interacting.  They have become a pair of vortex dipoles.  By $t=48$ they have completed their interaction and are moving away from each other around the torus.  The time slices in the figure are evenly spaced.  Thus it can be seen that, unlike in quasi-1D, the collision is not elastic.  The relative vortex velocity in the first three time slices, before the collision, is not the same as in the last two.  In fact the two vortex dipoles no longer have the same speed.

Although we have not pictured it here, the evolution of the above collision is very intriguing.  Despite the complicated density profile, it is apparent from the phase that the remnants of the two band solitons oscillate between single vortices, vortex dipoles, and band solitons.  The relationship between the spectra of these various collective excitations is an unsolved problem.

\subsection{Shock and phase rigidity}
\label{subsec:1Dshock}

Explorations of 2D and 3D out of the quasi-1D regime have thus far have been with periodic boundary conditions in the longitudinal direction.  If instead \emph{box} boundary conditions are used, together with a phase ramp of the kind desribed in the previous section, the BEC becomes chaotic.  As in quasi-1D, it exhibits the property of phase rigidity.  In response to shock it breaks up into well-defined phase domains.

To illustrate this effect a volume of 25x12 with box boundary conditions in both dimensions was used.  A large phase ramp of $\phi(x,y)=1.6\pi x + 0.2\pi y$ pushed a two-soliton, approximate stationary state into the walls at a skewed angle.  The time evolution of the resulting state is shown in figure~\ref{fig:shock2D}.  The first time slice shows the initial state.  The second time slice shows a shock wave.  By the final two time slices the condensate appears to have fractured into a chaotic assortment of phase domains.  The phase appears to be completely decoherent at scales larger than the healing length.

\section{Conclusion}
\label{sec:conclusion}

We have analytically described and numerically illustrated the experimental viability of soliton-train stationary states in quasi-one-dimension.  Soliton-trains can be manipulated with simple phase profiles.  In this way one can create a structured super-current or study soliton interactions.

In the repulsive BEC quasi-1D soliton-train solutions are stable.  In the attractive BEC those in which the members of the train are separated by nodes are stable.  Other solution-types are cyclically stable.  The attractive BEC is stable against collapse in the quasi-1D regime.

Beyond the quasi-1D regime we have shown that band and planar solitons decay into vortices when perturbed by white noise.  This is not only a new method of vortex creation but also allows for the direct study of vortex interactions.  We have also shown that the phase of the BEC is rigid; when shocked it shatters chaotically.

\ack
We benefited greatly from extensive discussions with Charles Clark and Nathan Kutz.  This work was supported by NSF Chemistry and Physics and the NSF Reseach Experience for Undergraduates program.

\section*{References}

\begin{thebibliography}{99}

\bibitem{davis1}
Davis K~B {\it et~al} 1995 {\it Phys. Rev. Letts.} {\bf 75} 3969

\bibitem{anderson1}
Anderson M~H {\it et~al} 1995 {\it Science} {\bf 269} 198  

\bibitem{fried1}
Fried D~G {\it et~al} 1998 {\it Phys. Rev. Letts.} {\bf 81} 3811  

\bibitem{bradley1}
Bradley C~C, Sackett C~A, Tollett J~J, and Hulet R~G 1995 {\it Phys. Rev. Letts.} {\bf 75} 1687  

\bibitem{bradley2}
Bradley C~C, Sackett C~A, and Hulet R~G 1997 {\it Phys. Rev. A} {\bf 55} 3951

\bibitem{dalfovo1}
Dalfovo F, Giorgini S, Pitaevskii L~P, and Stringari S 1999 {\it Rev Mod Phys} {\bf 71} 463

\bibitem{pitaevskii1}
Pitaevskii L~P 1961 {\it Sov. Phys. JETP} {\bf 13} 451

\bibitem{gross1}
Gross E~P 1961 {\it Nuovo Cimento} {\bf 20} 454  

\bibitem{kolomeisky1}
Kolomeisky E~B, Newman T~J, Straley J~P, and Qi X 2000  cond-mat/0002282

\bibitem{carr15}
Carr L~D, Clark C~W, and Reinhardt W~P 1999 {\it submitted to Phys. Rev. A e-print  cond-mat/9911177}

\bibitem{carr16}
Carr L~D, Clark C~W, and Reinhardt W~P 1999 {\it submitted to Phys. Rev. A e-print cond-mat/9911178}

\bibitem{ciao1}
Ciao R~Y, Deutsch I~H, Garrison J~C, and Wright E~W 1993 {\it Frontiers in
  Nonlinear Optics: the Serge Akhmanov Memorial Volume} (Bristol and Philadelphia: Institute of Physics Publishing) p 151

\bibitem{mamaev1}
Mamaev A~V, Saffman M, Anderson D~Z, and Zozuyla A~A 1996 {\it Phys. Rev. A} {\bf 54} 870

\bibitem{hasimoto1}
Hasimoto H 1972 {\it J. Fluid Mech.} {\bf 51} 477

\bibitem{hasegawa1}
Hasegawa A 1990 {\it Optical Solitons in Fibers} (New York: Springer-Verlag)

\bibitem{agrawal1}
Agrawal G~P 1995 {\it Nonlinear Fiber Optics 2nd ed} (San Diego: Academic Press)

\bibitem{deng1}
Deng L {\it et~al} 1999 {\it Nature} {\bf 398} 218

\bibitem{hagley1}
Hagley E~W {\it et~al} 1999 {\it Science} {\bf 283} 1706

\bibitem{mewes1}
Mewes M~O {\it et~al} 1997 {\it Phys. Rev. Letts.} {\bf 78} 582

\bibitem{key1}
Key M {\it et~al} 2000 {\it Phys. Rev. Letts.} {\bf 84} 1371

\bibitem{dekker1}
Dekker N~H {\it et~al} 2000 {\it Phys. Rev. Letts.} {\bf 84} 1124

\bibitem{andrews1}
 Andrews M~R {\it et~al} 1996 {\it Science} {\bf 273} 84

\bibitem{burger1}
Burger S {\it et~al} 1999 {\it e-print cond-mat/9910487}

\bibitem{ketterle1}
Ketterle W, Sturfee D~S, and Stamper-Kurn D~M 1999 {\it e-print cond-mat/9904034}

\bibitem{close1}
 Close J~D and Zhang W 1999 {\it J. Opt. B} {\bf 1} 420

\bibitem{matthews1}
Matthews M~R {\it et~al} 1999 {\it Phys. Rev. Letts.} {\bf 83} 2498

\bibitem{williams1}
Williams J~E and Holland M~J 1999 {\it Nature} {\bf 401} 568

\bibitem{carr19}
Carr L~D, Kutz N~T, and Reinhardt W~P 2000 {Phys. Rev. E to be submitted}

\bibitem{ruprecht1}
Ruprecht P~A, Holland M~J, Burnett K, and Edwards M 1995 {\it Phys. Rev. A} {\bf 51} 4704

\bibitem{kivshar5}
Kivshar Y~S and Alexander T~J 1999 {\it e-print cond-mat/9905048}

\bibitem{kivshar3}
Kivshar Y~S 1998 {\it Physics Reports} {\bf 298} 81

\bibitem{sulem1}
Sulem C and Sulem P~L 1999 {\it Nonlinear Schr\"odinger Equations: Self-focusing and Wave Collapse} (New York: Springer-Verlag)

\bibitem{zakharov1}
Zakharov V~E and Shabat A~B 1972 {\it Sov. Phys. JETP} {\bf 34} 62

\bibitem{zakharov2}
Zakharov V~E and Shabat A~B 1973 {\it Sov. Phys. JETP} {\bf 37} 823

\bibitem{miura1}
Miura R~M 1968 {J. Math. Phys.} {\bf 9} 1202

\bibitem{miura2}
Miura R~M, Gardner S~C, and Kruskal M~D 1968 {J. Math. Phys.} {\bf 9}

\bibitem{perez1}
P\'erez-Garc\'ia V~M, Michinel H, and Herrero H 1998 {\it Phys. Rev. A} {\bf 57} 3837

\bibitem{bowman1}
Bowman F 1961 {\it Introduction to Elliptic Functions, with Applications} (New York: Dover)

\bibitem{abramowitz1}
Abramowitz M and Stegun I~A (ed) 1964 {\it Handbook of Mathematical Functions} (Washington, D.C.: National Bureau of Standards)

\bibitem{reinhardt1}
Reinhardt W~P and Clark C~W 1997 {\it J. Phys. B} {\bf 30} L785

\bibitem{denschlag1}
Denschlag J and {\it et al} 2000 {\it Science} {\bf 287} 97

\bibitem{kivshar1}
Kivshar Y~S and Yang X 1994 {\it Phys. Rev. E} {\bf 49} 1657

\bibitem{elgin1}
Elgin J~N 1993 {\it Phys. Rev. A} {\bf 47} 4331

\bibitem{berge1}
Berge L, Alexander T~J, and Kivshar Y~S 1999 {\it e-print cond-mat/9907408}

\bibitem{jones1}
Jones C~A, Putterman S~J, and Roberts P~H 1986 {\it J. Phys. A: Math. Gen.} {\bf 19} 2991

\bibitem{michinel1}
Michinel H, P\'erez-Garc\'ia V~M, and de~la Fuente R 1999 {\it Phys. Rev. A} {\bf 60} 1513

\bibitem{dobrek1}
Dobrek L {\it et~al} 1999 {\it Phys. Rev. A} {\bf 60} R3381

\bibitem{anderson2}
Anderson P~W 1984 {\it Basic Notions of Condensed Matter Physics} (New York: Addison-Wesley)

\endbib

\begin{table}
\caption{Physical time and length scales of our numerical studies.  Thus 100 time units is 20-200 ms.  Note that the number of atoms scales with the trap volume, so that $10^4$ atoms would be found in the quasi-1D regime of current experimental volumes.  We fix the healing length at 3 $\mu$m for the sake of observability.}
\lineup
\begin{indented}
\item[]\begin{tabular}{@{}llllll}
\br
atom   &V ($\xi_{eff}^3$)&$\xi_{eff}$ ($\mu$m)&$t$ (ms)&$N$     &$a_s$ (nm)\\
\mr
$^{23}$Na&1x1x25     &3.0           &0.727   &\01,610   &\02.75\\
$^{87}$Rb&1x1x25     &3.0           &2.75    &\0\0$\,$769   &\05.77\\
$^{1}$H  &1x1x25     &3.0           &0.0316  &76,500   &\00.0581\\
$^7$Li &1x1x10       &3.0           &0.221   &\01,250   &\0\-1.45\\
\br
\end{tabular}
\end{indented}
\label{tbl:exp}
\end{table}

\begin{figure}
\begin{center}
\epsfig{file=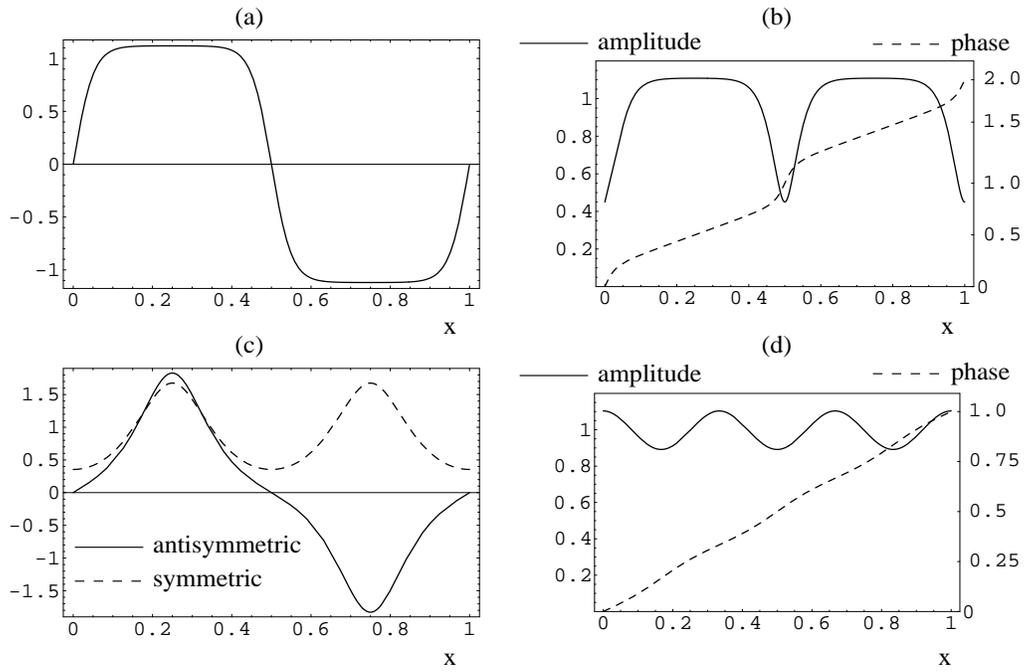,height=3.5in,width=5.5in} 
\end{center}
\caption{Shown are the five stationary solution-types in one dimension under periodic boundary conditions.  Repulsive case: (a) amplitude of real solution and (b) amplitude and phase of complex solution.  Attractive case: (c) amplitude of real symmetric and antisymmetric solutions and (d) amplitude and phase of complex solution.  All plots are normalized to the unit interval.}
\label{fig:5solns}
\end{figure}

\begin{figure}
\begin{center}
\epsfig{file=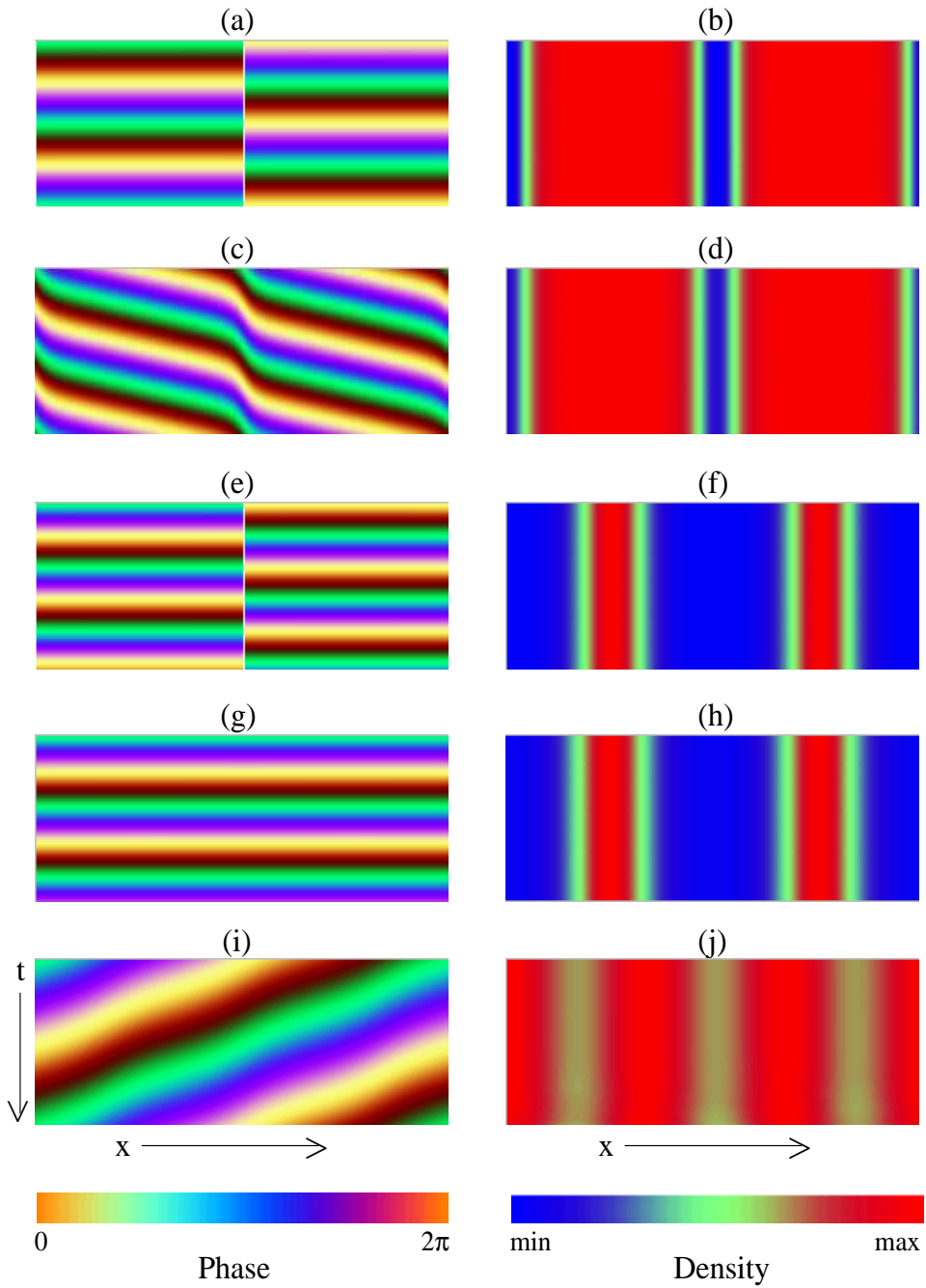,height=7in,width=5in} 
\end{center}
\caption{The five stationary solution-types in one dimension under periodic boundary conditions are propagated in time numerically for 100 natural time units.  Shown are phase and density for (a)-(b) real solutions, repulsive case; (c)-(d) complex solutions, repulsive case; (e)-(f) anti-symmetric real solutions, attractive case; (g)-(h) symmetric real solutions, attractive case; (i)-(j) complex solutions, attractive case.}
\label{fig:1D}
\end{figure}

\begin{figure}[t]
\begin{center}
\epsfig{file=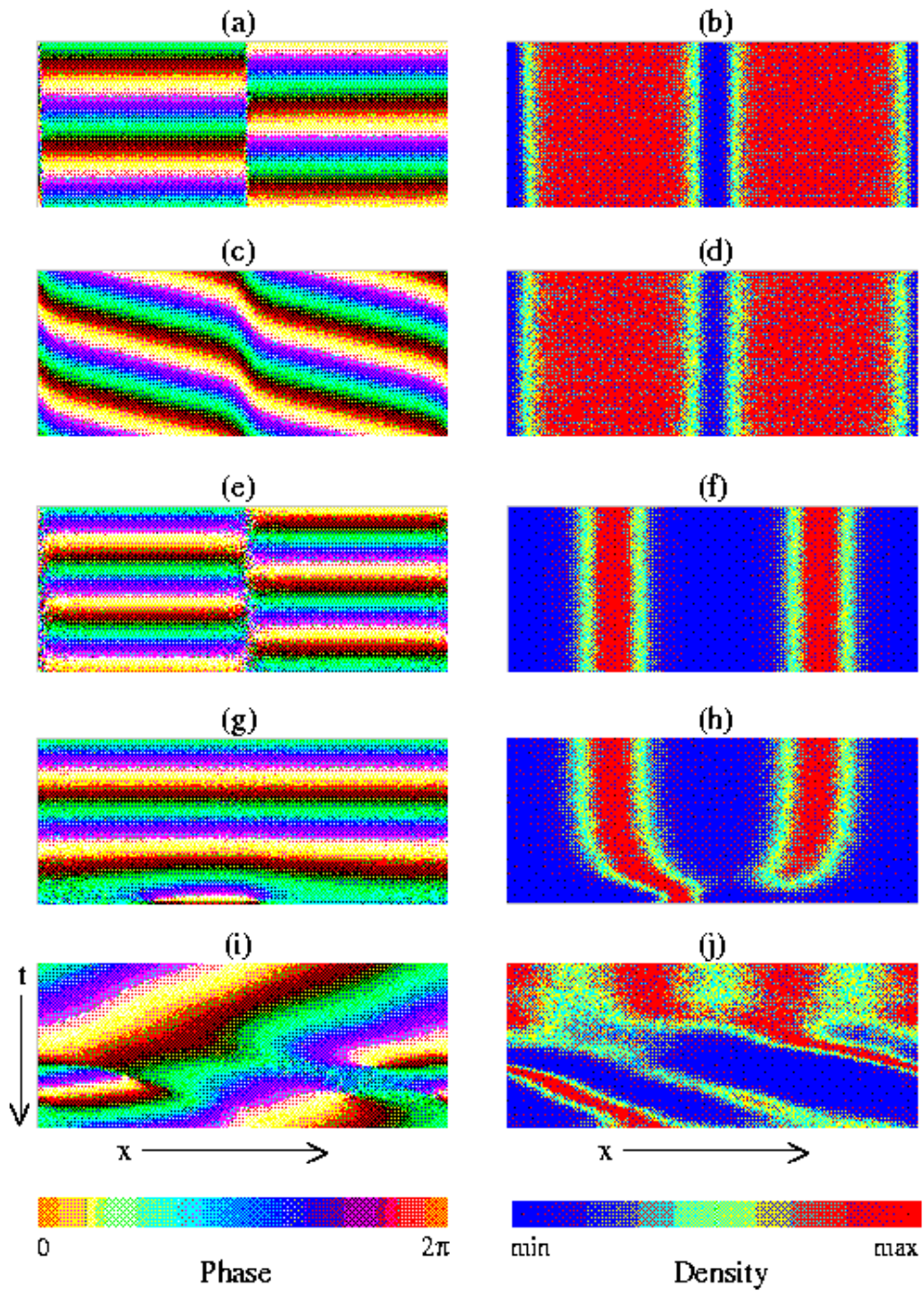,height=7in,width=5in} 
\end{center}
\caption{The five stationary solution-types in one dimension under periodic boundary conditions are propagated in time numerically, as in figure~\ref{fig:1D}, for 100 natural time units.  Stochastic noise at the level of $1\%$ is added to the initial state.  (a)-(f) are stable; (g)-(j) are cyclically stable.  Shown are phase and density for (a)-(b) real solutions, repulsive case; (c)-(d) complex solutions, repulsive case; (e)-(f) anti-symmetric real solutions, attractive case; (g)-(h) symmetric real solutions, attractive case; (i)-(j) complex solutions, attractive case.}
\label{fig:1Dnoise}
\end{figure}

\begin{figure}
\begin{center}
\epsfig{file=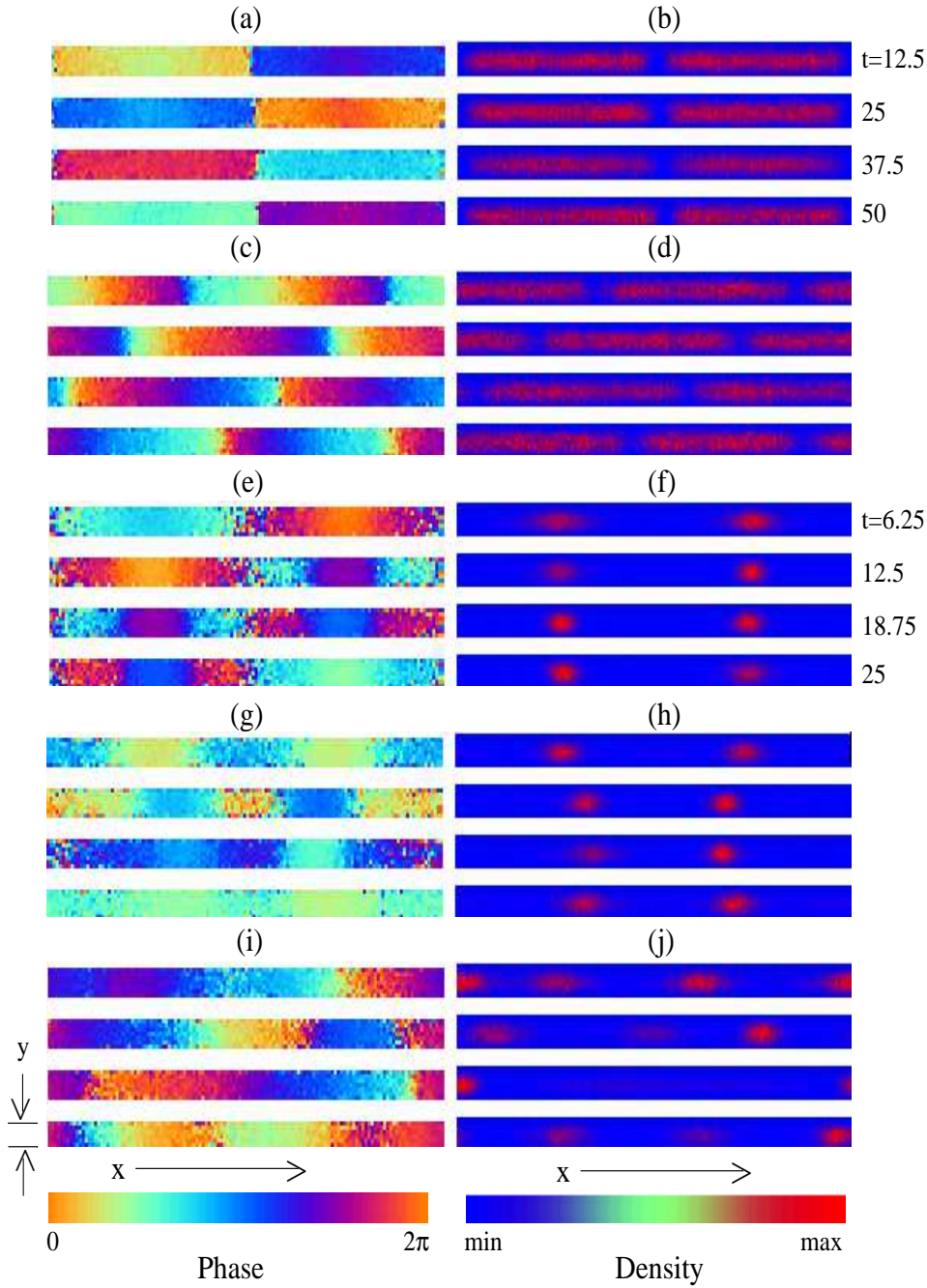,height=7in,width=5in} 
\end{center}
\caption{The five stationary solution-types are extended into quasi-one-dimension under periodic boundary conditions.  They are propagated in time numerically as in figure~\ref{fig:1Dnoise}, for 50 and 25 natural time units in the repulsive and attractive cases.  Stochastic noise at the level of $1\%$ is added to the initial state.  (a)-(f) are stable; (g)-(j) are cyclically stable.  Shown are four time slices through the $xy$ plane of phase and density of fully 3D simulations for (a)-(b) real solutions, repulsive case; (c)-(d) complex solutions, repulsive case; (e)-(f) anti-symmetric real solutions, attractive case; (g)-(h) symmetric real solutions, attractive case; (i)-(j) complex solutions, attractive case.}
\label{fig:quasi1D}
\end{figure}

\begin{figure}
\begin{center}
\epsfig{file=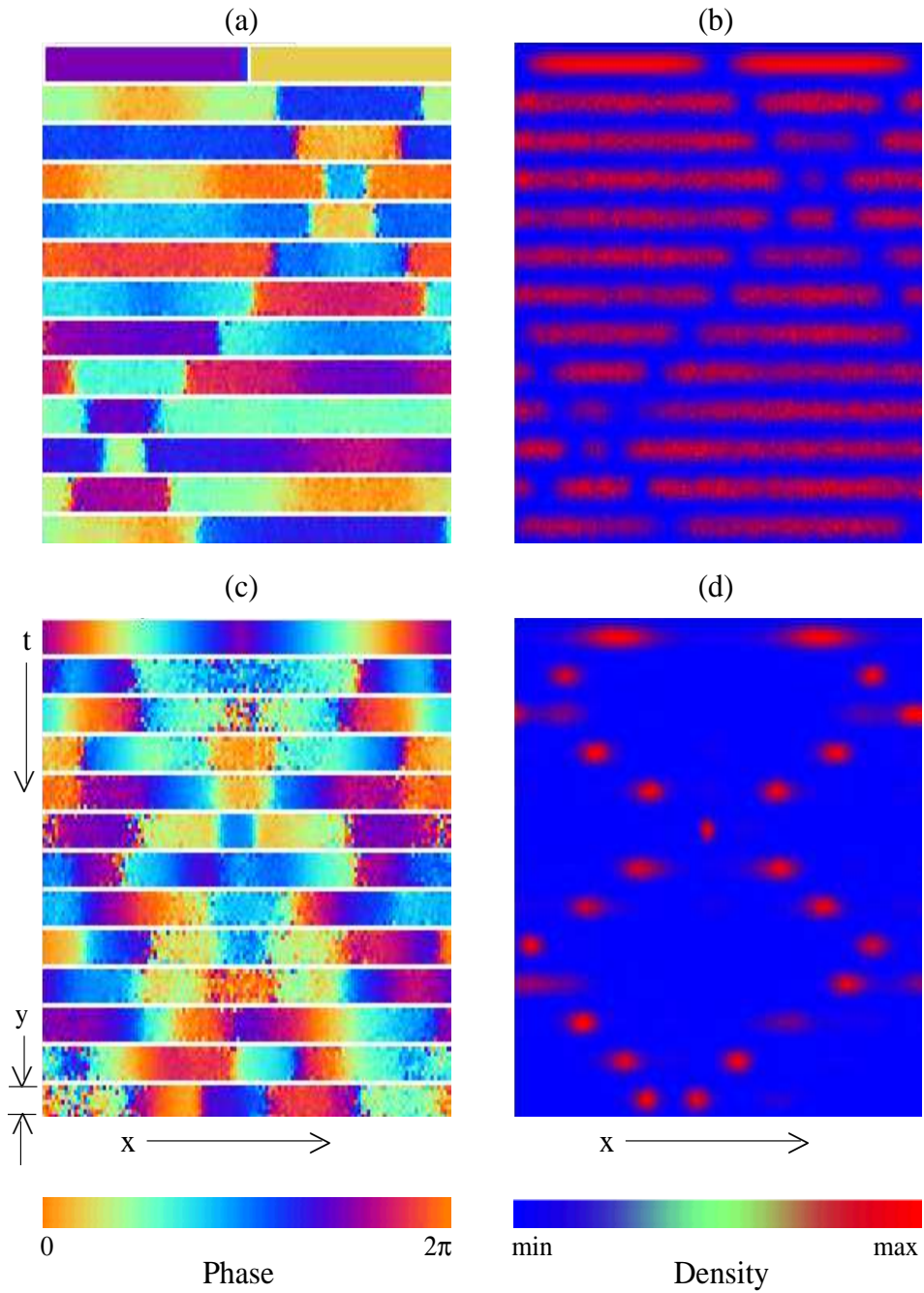,height=7in,width=5in} 
\end{center}
\caption{In quasi-1D soliton-train solutions can be manipulated by simple phase profiles.  Here we show examples of such manipulation for (a)-(b) the repulsive and (c)-(d) the attractive cases.  The solitons have been given equal and opposite velocity.  The evolution of their phase and density is shown by plotting the $xy$ plane at the midpoint of the $z$ co-ordinate in time slices running down the page.  We have added $1\%$ stochastic noise to the initial state.  The results shown follow from a fully 3D propagation with dimensions 25x1x1 in the repulsive case and 10x$\frac{1}{2}$x$\frac{1}{2}$ in the attractive case.}
\label{fig:phaseManip}
\end{figure}

\begin{figure}
\begin{center}
\epsfig{file=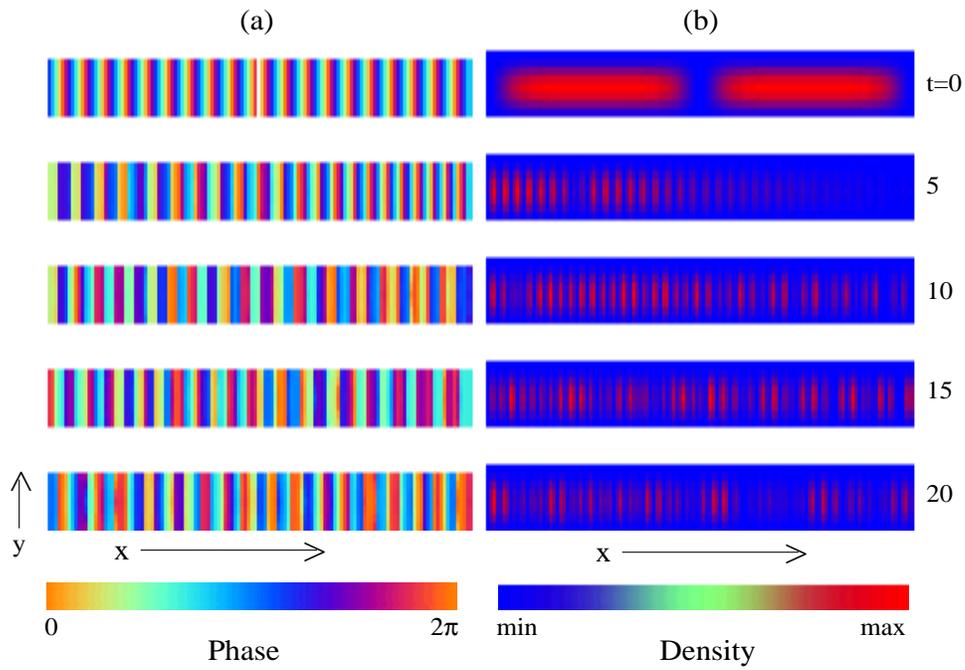,height=3.5in,width=5in} 
\end{center}
\caption{A rapidly translating condensate in a quasi-1D box is suddenly stopped.  The condensate \emph{shatters} into domains of constant phase due to the fragility of this phase-symmetry-breaking system.}
\label{fig:shock1D}
\end{figure}

\begin{figure}
\begin{center}
\epsfig{file=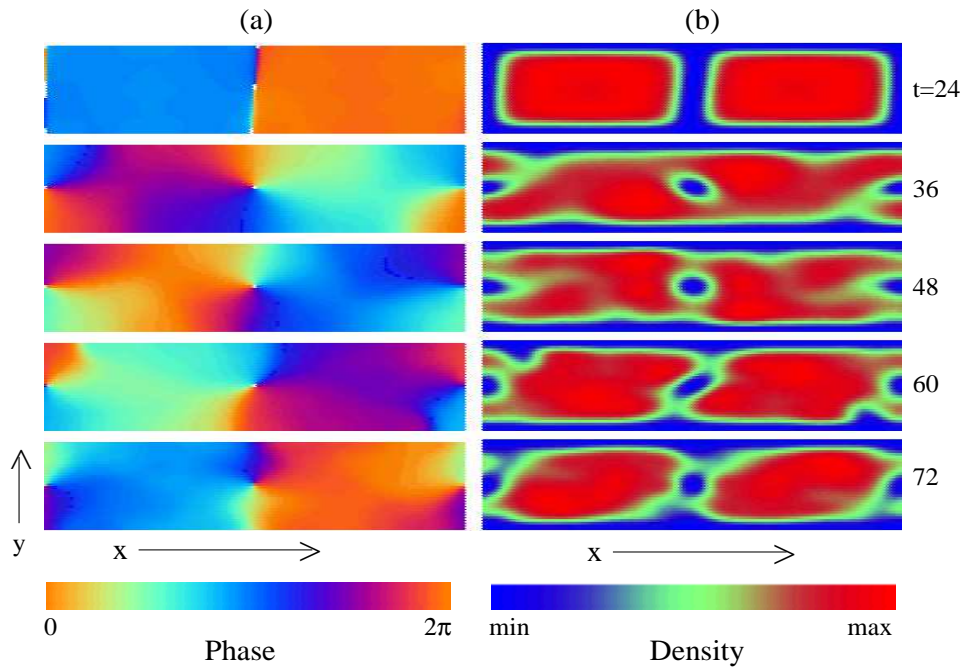,height=3.5in,width=5in} 
\end{center}
\caption{In 2D band solitons decay into pairs of oppositely charged vortices.  We have added $0.01\%$ stochastic noise to an initial two-soliton stationary solution in order to demonstrate this effect.  Shown are five slices of the $xy$ plane at the midpoint of the $z$ co-ordinate, equally spaced in time.  The longitudinal direction has periodic boundary conditions.  Note that the phase leads the instability.}
\label{fig:vortex2D}
\end{figure}

\begin{figure}
\begin{center}
\epsfig{file=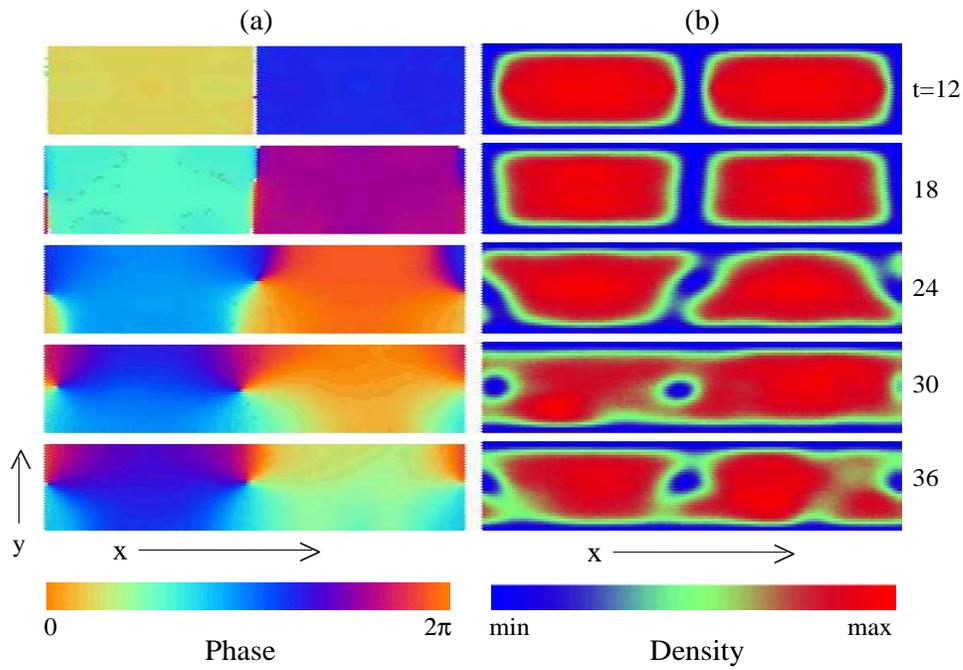,height=3.5in,width=5in} 
\end{center}
\caption{In 3D planar solitons decay via a snake instability.  We have used the same noise level as in figure~\ref{fig:vortex2D}.  We have highlighted an earlier set of times, as the decay occurs earlier than in 2D.  Shown is the $xy$ plane at the midpoint of the $z$ co-ordinate.}
\label{fig:vortex3D}
\end{figure}

\begin{figure}[t]
\begin{center}
\epsfig{file=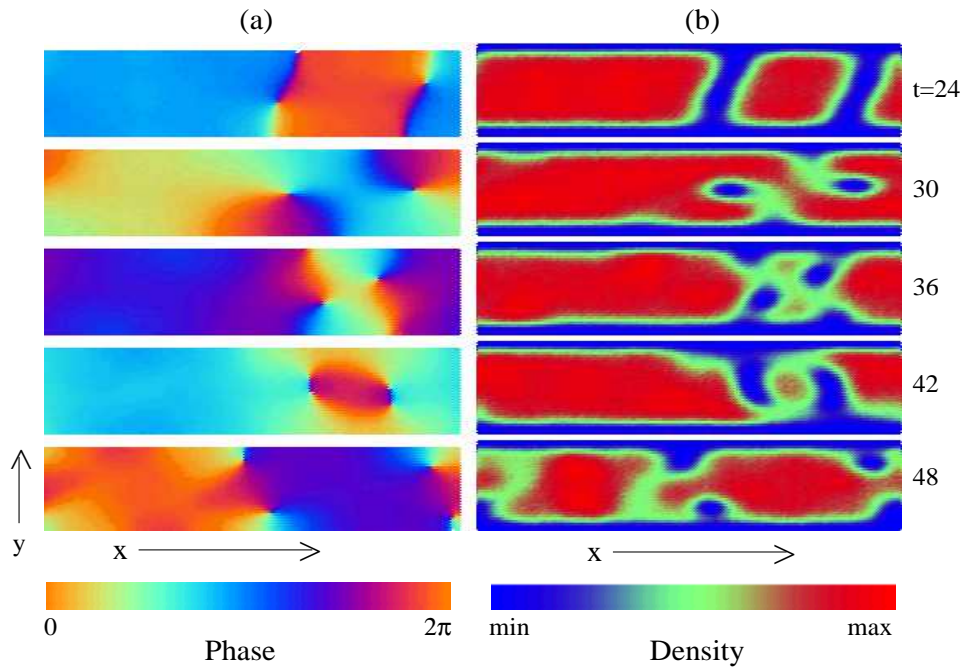,height=3.5in,width=5in}
\end{center}
\caption{Collision of two initial band solitons which have evolved into two oppositely charged vortex pairs.  We have added $0.1\%$ stochastic noise.  As in figure~\ref{fig:phaseManip} the relative velocity is induced by phase discontinuities across the band solitons, but the interaction is that of two vortex dipoles.  This is an example of how to use the noise-induced instability of band solitons in 2 and 3D to study vortex interactions.}
\label{fig:interact}
\end{figure}

\begin{figure}
\begin{center}
\epsfig{file=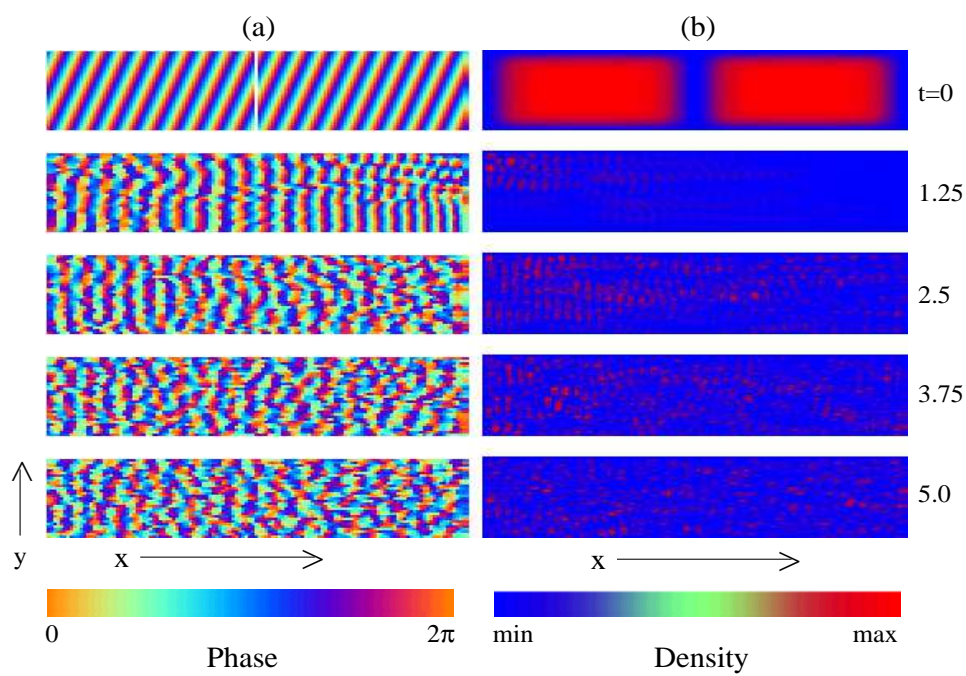,height=3.5in,width=5in} 
\end{center}
\caption{A rapidly translating quasi-2D repulsive condensate is stopped.  Following propagation of a density-dependent shock front, a fully chaotic phase and spatial distribution is obtained.}
\label{fig:shock2D}
\end{figure}

\end{document}